\documentclass[]{spie}  %>>> use for US letter paper

\usepackage{amsmath,amsfonts,amssymb,mathrsfs}
\usepackage{graphicx}
\usepackage[colorlinks=true, allcolors=blue]{hyperref}
\usepackage{astro_bib_macro} %shortcuts of journals defined in this file
\usepackage{authblk}
\usepackage{float} % to force figure position with \begin{figure}[H]

\title{Visible extreme adaptive optics on extremely large telescopes:\\Towards detecting oxygen in Proxima Centauri b and analogs}

\author[a]{J. Fowler}
\affil[a]{Department of Astronomy \& Astrophysics, University of California, Santa Cruz, CA, USA}
\author[b]{Sebastiaan Y. Haffert}
\affil[b]{Steward Observatory, University of Arizona, Tucson, AZ, USA}
\author[c]{Maaike A.M. van Kooten}
\affil[c]{National Research Council Canada, Herzberg Astronomy and Astrophysics Research Center, Victoria, Canada}
\author[d]{Rico Landman}
\affil[d]{Leiden Observatory, Leiden University, Leiden, The Netherlands}
\author[e]{Alexis Bidot}
\author[e]{Adrien Hours}
\affil[e]{Univ. Grenoble Alpes, CNRS, IPAG, Grenoble, France}
\author[f]{Mamadou N’Diaye}
\affil[f]{Universit\'e C\^ote d’Azur, Observatoire de la C\^ote d’Azur, CNRS, Laboratoire Lagrange, France}
\author[g]{Olivier Absil}
\affil[g]{STAR Institute, Universit\'e de Li\`ege, Li\`ege, Belgium}
\author[h]{Lisa Altinier}
\affil[h]{Aix Marseille Univ, CNRS, CNES, LAM, Marseille, France}
\author[i]{Pierre Baudoz}
\affil[i]{LESIA, Observatoire de Paris, Universit\'e PSL, CNRS, Sorbonne Universit\'e, Universit\'e de Paris, Meudon, France}
\author[f]{Ruslan Belikov}
\affil[f]{NASA Ames Research Center, Moffett Field, USA}
\author[g, h]{Markus Johannes Bonse}
\affil[g]{Max Planck Institute for Intelligent Systems, Tübingen, Germany}
\affil[h]{Department of Physics, ETH Zurich, Switzerland}
\author[i, j]{Kimberly Bott}
\affil[i]{University of California, Riverside}
\affil[j]{NASA Nexus for Exoplanet System Science, Virtual Planetary Lab Team, Seattle, USA}
\author[d]{Bernhard Brandl}
\author[e]{Alexis Carlotti}
\author[k]{Sarah L. Casewell}
\affil[k]{School of Physics and Astronomy, University of Leicester, Leicester, UK}
\author[h]{Elodie Choquet}
\author[l]{Nicolas B. Cowan}
\affil[l]{Department of Earth \& Planetary Sciences and Department of Physics, McGill University, Montr\'eal, QC, Canada}
\author[m]{Niyati Desai}
\affil[m]{Department of Astronomy, California Institute of Technology, Pasadena, CA, USA}
\author[d,n]{David Doelman}
\affil[n]{SRON Netherlands Institute for Space Research, Leiden, The Netherlands}
\author[f]{Kevin Fogarty}
\author[g,h]{Timothy D. Gebhard}
\author[i,o,p]{Yann Gutierrez}
\affil[o]{DTIS, ONERA, Universit\'e Paris Saclay, Palaiseau, France}
\affil[p]{DOTA, ONERA, Ch\^atillon, France}
\author[b,q,r,s]{Olivier Guyon}
\affil[q]{Subaru Telescope, NAOJ, USA}
\affil[r]{College of Optical Sciences, University of Arizona, Tucson, AZ, USA}
\affil[s]{Astrobiology Center, Osawa, Mitaka, Tokyo, Japan}
\author[o]{Olivier Herscovici-Schiller}
\author[t,u]{Roser Juanola-Parramon}
\affil[t]{NASA Goddard Space Flight Center, Greenbelt, MD, USA}
\affil[u]{University of Maryland Baltimore County, 1000 Hilltop Cir, Baltimore, MD, USA}
\author[d]{Matthew Kenworthy}
\author[d]{Elina Kleisioti}
\author[g]{Lorenzo K\"onig}
\author[v]{Mariya Krasteva}
\affil[v]{European Space Agency, ESTEC, The Netherlands}
\author[i]{Iva Laginja}
\author[e]{Lucie Leboulleux}
\author[g]{Johan Mazoyer}
\author[w]{Maxwell A. Millar-Blanchaer}
\affil[w]{Department of Physics, University of California, Santa Barbara, CA, USA}
\author[e]{David Mouillet}
\author[x]{Emiel Por}
\author[x]{Laurent Pueyo}
\affil[x]{Space Telescope Science Institute, Baltimore, MD, USA}
\author[d]{Frans Snik}
\author[d]{Dirk van Dam}
\author[b]{Kyle van Gorkom}
\author[y]{Sophia R. Vaughan}
\affil[y]{University of Oxford, UK}
\begin{document} 
\maketitle

\begin{abstract}
 Looking to the future of exo-Earth imaging from the ground, core technology developments are required in visible extreme adaptive optics (ExAO) to enable the observation of atmospheric features such as oxygen on rocky planets in visible light. UNDERGROUND (Ultra-fast AO techNology Determination for Exoplanet imageRs from the GROUND), a collaboration built in Feb. 2023 at the Optimal Exoplanet Imagers Lorentz Workshop, aims to (1) motivate oxygen detection in Proxima Centauri b and analogs as an informative science case for high-contrast imaging and direct spectroscopy, (2) overview the state of the field with respect to visible exoplanet imagers, and (3) set the instrumental requirements to achieve this goal and identify what key technologies require further development. 
\end{abstract}


%\section*{sources}

%\href{https://docs.google.com/presentation/d/1Ot4Yahrr0MxqgqUGkuHceXUiVsR9HnZmhRjs0yRNvuY/edit#slide=id.g21121a3c60d_3_16}{Google slides}, \href{https://docs.google.com/document/d/10lswqczfTDfzE0B9ki2DehwOd5AIlOAbikzpx2f-e3w/edit}{Google doc}, \href{https://app.slack.com/client/T04NNJ1GENS/C04RE2LUK0Q}{Slack channel}

%%%%%%%%%%%%%%%%%%%%%%%%%%%%%%%%%%%%%%%%%%%%%%%%%%%%%%%%%%%%%%%
\section{Introduction} \label{sec:intro} % This includes the problem statement? 
% Hunting for oxygen in the visible for Proxima B
%%%%%%%%%%%%%%%%%%%%%%%%%%%%%%%%%%%%%%%%%%%%%%%%%%%%%%%%%%%%%%%

Finding and characterizing exo-Earths is a key science goal of Extremely Large Telescopes, specifically to explore the diversity of substellar companions, understand the formation and evolution of planetary systems, and find clues about the presence of life outside our solar system. Unlike their gas giant-planet cousins, Earth analogs will have an extremely faint self-luminous glow in the near infrared (NIR); current and future ground-based visible to NIR exoplanet imagers will not be sensitive enough to detect that thermal emission. To observe these rocky planets we must detect them in reflected light from their host star.

Proxima Centauri is a promising stellar system for reflected light studies and famous for its proximity to Earth. Proxima Centauri A is an M dwarf at 1.29 pc from our solar system that hosts the Earth-like planet Proxima Centauri b with a mass of 1.3 $M_{\oplus}$. Our closeness to Proxima Centauri is favorable to study this system with coronagraphic observations, as planetary companions appear at larger angular separations from their host star as they get closer to Earth. For example, Proxima Centauri b has a maximum separation from its host star of 38 milli-arcseconds (mas) \cite{proxima2016}. Such an orbit is in theory resolvable at visible wavelengths (e.g. at $\lambda_{O_2}$=765\,nm) by current telescopes such as the VLT or Magellan ($\lambda_{O_2}/D$ = 19 mas, and $\lambda_{O_2}/D$ = 24 mas, respectively) and some initiatives exist to study these features in the Alpha Centauri system\cite{kasper2017}, but the the light gathering efficiency of an 8-10 meter telescope makes high resolution spectroscopy extremely challenging. These observations become more feasible with the forthcoming extremely large observatories, with a resolution of 4 and 6 mas for the European Extremely Large Telescope (ELT) and the Giant Magellan Telescope (GMT).

While many spectral features indicate the disequlibrium chemistry that implies signs of life, oxygen is thought to be one of the strongest biosignatures, as it is the strongest marker of life in Earth's current atmosphere. The abundance of O$_2$ closely follows the presence of life that uses oxygenic photosynthesis mechanisms. While recent research has found that there are also several abiotic processes that can generate an O$_2$ signature, we still require the detection of O$_2$ as a prerequisite for the presence of life \cite{meadows2018exoplanet}. In this work we examine the requirements to detect oxygen in the atmosphere of Proxima Centauri b using ground-based telescopes. 

Section \ref{sec:hci_from_ground} describes the current state of the field for visible light adaptive optics from the ground. In Section \ref{sec:contrast_requirements} we outline the required contrast to detect oxygen in the atmosphere of Proxima Centauri b. In Section \ref{sec:performance} we estimate a simplified AO wavefront error budget to determine the speed of the AO system before simulating the final contrast in Section~\ref{sec:ContrastCurves}. Section \ref{sec:underground} suggests an instrument architecture and describes the remaining technology development required to reach it. Finally, in Section \ref{sec:conc} we summarize this work and make our final conclusions. 

\section{State of Visible High-Contrast Imaging from the Ground}\label{sec:hci_from_ground}

Only a handful of extreme Adaptive Optics (ExAO) instruments currently operate at visible wavelengths: VLT/SPHERE \cite{beuzit2019sphere}, SUBARU/SCExAO \cite{jovanovic2015subaru} and Magellan/MagAO-X \cite{males2018magaox, close2018optical}. These facilities report raw contrasts on the order of $\sim10^{-3}$ at 100 mas, see Table \ref{tab:contrast_limits}. The leading limitation in all systems are non common path aberrations (NCPA) which can be both static and quasi-static. To reduce the impact of these errors, one of the most crucial developments is active focal plane wavefront control, especially the algorithms that can control mid to high-spatial frequency wavefront errors. Electric Field Conjugation (EFC) and implicit-EFC (iEFC) are examples of such dark hole digging algorithms \cite{potier2022,haffert2023,ahn2023}. These methods have been proven to remove quasi-static aberrations to within $5\times 10^{-8}$ in the lab \cite{haffert2023} and $1\times 10^{-6}$ on sky \cite{potier2022}. After the NPCA speckles have been removed, atmospheric speckles are the ultimate limiting factor on raw contrasts.

\begin{table}[h!]
\begin{center}
\caption{ State of the art for visible AO performance on 8m-class telescopes. }
\begin{tabular}{l|l|l|l|l}
\hline
Instrument                            & Contrast & Wavelength/Band & Seeing & Ref. \\ \hline \hline
\multicolumn{1}{l|}{SPHERE/ZIMPOL}   & $10^{-3}$     & I'   & 0.9" &  \cite{schmid2018sphere} \\ \hline 
\multicolumn{1}{l|}{SCExAO/VAMPIRES} & 5$\times 10^{-3}$     & 750 nm     &   0.55"\textsuperscript{*}   &  \cite{miles2022} \\ \hline
\multicolumn{1}{l|}{MagAO-X}         & 2$\times 10^{-3}$     & I      & 0.75" &  private communication \\ \hline
\end{tabular}
\label{tab:contrast_limits}
\end{center}
\small\textsuperscript{*} As the seeing for this measurement was not recorded, we have provided the median seeing for the Mauna Kea Summit.
\end{table}

\section{Contrast requirements for detecting oxygen}

\subsection{Gain from high-dispersion spectroscopy}
%High-dispersion spectroscopy (HDS) can very efficiently filter out starlight. 
An exoplanet atmosphere is notably different from that of its host star due to a difference in atomic and molecular abundances and pressure-temperature profiles. At a high enough spectral resolution, this can be used to disentangle planet and starlight by using matched filters \cite{snellen2015combining}, making high-dispersion spectroscopy (HDS) an efficient way to filter out starlight. However, using HDS leaves only the information content that is present in the spectral lines; there is a loss of all continuum light after filtering since the majority of the planet continuum is not different from that of its host star. 

At low spectral resolution it is not possible to disentangle starlight from planet light, as stellar speckles and planets have a similar spectral shape \cite{landman2023trade}. In this case, a low-pass filter on the measured spectra is required to remove the influence of star light which will also remove almost all of the continuum flux of the planet. 

\begin{figure}[h!]
\makebox[\textwidth][c]{
    \mbox{\includegraphics[width=0.6\textwidth]{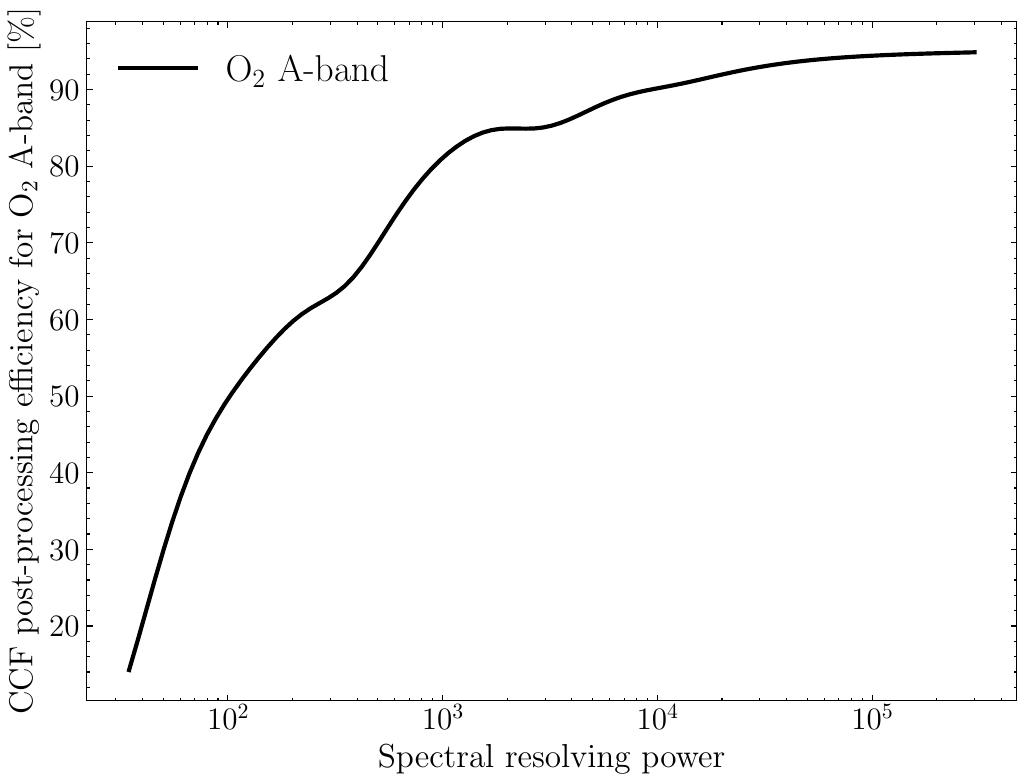}}
    }
    \caption{Efficiency of HDS post-processing as a function of resolving power. In this case we use a cross correlation function. The efficiency converges to 95\% around $R=10^5$.}
\label{fig:igc}
\end{figure}

The efficiency of HDS is defined as the ratio between the signal after and before low-pass filtering \cite{landman2023trade}. This efficiency for the oxygen A-band lines is shown in Figure \ref{fig:igc}, calculated with a spectral bandwidth ranging between 760 and 770 nm, corresponding to the range for the oxygen A-band. For our observing case, we assume observations of the system in the most optimistic time window, in which the influence of Earth's telluric lines are minimized \cite{rodler2014feasibility, serindag2019testing}. The plot echoes previous results\cite{rodler2014feasibility, serindag2019testing, hardegree2023bioverse} that the efficiency is highest at high spectral resolution.

\label{sec:contrast_requirements}
\subsection{Contrast requirements}
Our science case is the detection of the oxygen A-band line in the atmosphere of Proxima Centauri b (wavelength centered at 765nm; 10nm bandwidth; 10$^{-7}$ contrast) within a single night of observation with the ELT. %High-spectral resolution (R$=2\times 10^5$) observations are required to separate the oxygen lines from Earth's telluric lines. 
The instrument will need simultaneous spectra of the star and planet for removing residual star light. Since HDS post processing algorithm is a photon-noise limited method, SNR is calculated with
\begin{equation}
\mathrm{SNR} = \frac{\eta T_p C F_{s}}{\sqrt{F_s K}}=\eta C T_p \sqrt{F_s/K},
\end{equation}
with efficiency of the post-processing $\eta$, throughput of the planet $T_p$, contrast between planet and star $C$, stellar flux $F_s$, and achieved raw contrast $K$. The main drivers for the SNR are $\eta$, $T_p$, and $K$. We outline the parameters for our calculation in Table \ref{tab:contrast_parameters}. This leads to a raw contrast requirement of $3\times 10^{-5}$. 

\begin{table}[h!]
\begin{center}
\caption{Parameters used to calculate the required raw contrast for Proxima Centauri b.  }
\begin{tabular}{l|l}
\hline
parameter  & value used for simulations \\ \hline \hline
 ELT telescope area & 1058.32 m\\ \hline 
 Spectrograph throughput      & 2 0.1 (end-to-end)   \\ \hline
 $\eta$ (HDS observing efficiency)     & 0.95 (from Figure \ref{fig:igc}   \\ \hline
 Spectral bandwidth  & 10 nm (760 nm t0 770 nm)  \\ \hline
 Contrast of Proxima Centauri b & $1\times 10^{-7}$ \\ \hline
 Stellar magnitude of Proxima Centauri A & 7.4 mag in I band \\ \hline
 integration time & 4 hours \\ \hline
 observing goal      & SNR=5   \\ \hline
\end{tabular}
\label{tab:contrast_parameters}
\end{center}
\end{table}

%\begin{itemize}
%\item ELT with a telescope area of 1058.32 m$^2$
%\item Contrast of Proxima Centauri b at $1\times 10^{-7}$
%\item End-to-end throughput to spectrograph of 0.1
%\item SNR goal of 5
%\item HDS efficiency $\eta$ of 0.95, according to Figure \ref{fig:igc}.
%\item a spectral bandwidth of 10.0 nm (760 nm to 770 nm)
%%\item Proxima Centauri with a stellar magnitude of 7.4 at I-band
%\item a total integration time of 4 hours
%\end{itemize}
%With these parameters we achieve a raw contrast requirement of $3\times 10^{-5}$ at 10 $\lambda/D$ to achieve a SNR of 5 in 4 hours of exposure time with HRS on the ELT.

\label{sec:performance}
\subsection{Adaptive optics error budget}

We calculate a simplified AO wavefront error budget to investigate the effects of the time delay in the AO system (i.e., the servo-lag error) on the wavefront error and Strehl ratio. From our contrast requirement in Section~\ref{sec:performance}, we can estimate the required wavefront error (contrast is inversely proportional to the square of the wavefront error in radians for a given separation). Over the entire control radius we need to maintain a wavefront error around 45nm RMS.  

Using the analytical AO modelling software \texttt{TIPTOP}~\cite{neichel2021tiptop}, we determine the fitting error for a system using a deformable mirror (DM) with 200 actuators across the pupil diameter. Here we make the choice of using this DM actuator density to match the actuator pitch roughly to the Fried parameter $r_0$ expected for good seeing conditions on the ELT ($r_0=$ 20 cm). We then look at the effect of the servo-lag error on the total residual wavefront error and use that to estimate the Strehl ratio (SR) using the Marechal approximation. The largest error term is the fitting error which will be reduced for the best seeing conditions or with future technology that further increases the actuator density on wavefront correctors such as a DM. Assuming the wavefront sensor aliasing is negligible by using an optimal wavefront sensor, the next largest error term is the servo-lag error. 

\begin{table}[h!]
\begin{center}
\caption{RMS (root mean square) wavefront error in nm for major error terms as we vary the control speed. Assuming a 39m telescope, natural guide star (not limited by SNR/guidestar magnitude), 0.5" seeing, 10 m/s effective wind velocity, 200 actuators across the pupil diameter, and loop delay of 1 frame, at 750 nm.}
\begin{tabular}{l|l|l|l}
\hline
Error terms at different control loop speed                          & 1kHz & 2kHz & 5kHz \\ \hline \hline
\multicolumn{1}{l|}{Fitting Error (nm RMS)}   & 38.92     & 38.92    & 38.92 \\ \hline
\multicolumn{1}{l|}{Servo-lag Error (nm RMS)}   &   29.47  &   14.87 &  5.97       \\ \hline
\multicolumn{1}{l|}{Total residual wavefront Error (nm RMS)}  &  48.8     & 41.7     &   39.38    \\ \hline\hline
\multicolumn{1}{l|}{Strehl Ratio (Marechal approx.)}   & 85\%     & 89\%    &   90\%   \\ \hline
\end{tabular}
\label{tab:wavefront_error_budget}
\end{center}
\end{table}

In Table~\ref{tab:wavefront_error_budget}, we present the effects of these two terms while changing the speed of the AO system, effectively tuning the servo-lag error. We are in a regime where reducing the servo-lag error provides an important improvement in SR allowing up to 90\% SR when running at 5kHz. This shows that we need to run at approximately 2kHz to achieve a wavefront error smaller than 45\,nm RMS and deliver images with high Strehl ratio to the coronagraph. While minimizing the wavefront error before the coronagraph is important to achieving good contrast, we need to build on these simulations and look directly at the exact impact of the servo-lag error and our proposed system on the raw contrast; these results are shown in Section~\ref{sec:ContrastCurves}. 

\subsection{Contrast curves at varying AO loop speeds}\label{sec:ContrastCurves}
A more detailed analysis of the performance was done by using a spatial-filter based semi-analytical approach to model the AO performance \cite{jolissaint2010synthetic, males2018ground}. The semi-analytical approach models each aspect of the AO system as a spatial transfer function that acts onto the atmospheric Power Spectral Densities (PSDs). The final PSD is then propagated through coronagraph to get the achieved raw contrast. This is also done by using a semi-analytical approach, making this method an efficient way to generate long-exposure PSFs \cite{herscovici2017analytic}. The noise on each mode is \cite{chambouleyron2021},
\begin{equation}
    \sigma^2 = \frac{1}{s_{pn}^2 N} + \frac{N_{sub}\sigma_{rn}^2}{s_{rn}^2 N^2}.
\end{equation}
with $\sigma^2$ the reconstructed wavefront variance, $s_{pn}$ the sensitivity to photon noise, $N_{sub}$ the number of pixels used for wavefront sensing, $\sigma_{rn}$ the detector read noise, $s_{rn}$ the read noise sensitivity and $N$ the number of photons per frame. 

\begin{figure}[h!]
\makebox[\textwidth][c]{
\mbox{\includegraphics[width=0.7\textwidth]{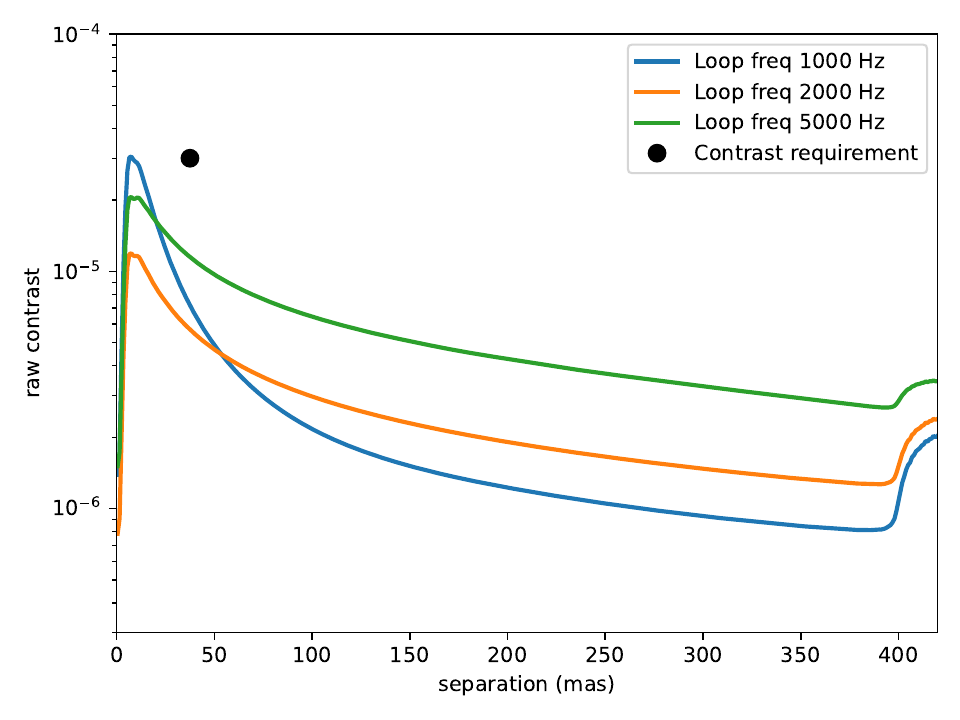}}
}
    \caption{Raw contrast as a function of angular separation for various (1, 2, and 5 kHz) loop speeds with an unmodulated pyramid wavefront sensor (PWFS). The black dot illustrates the contrast requirement for the detection of oxygen on Proxima Centauri b. Note that 2kHz AO provides the best correction at small angular separations.}
\label{fig:contrast_curves}
\end{figure}

The pyramid wavefront sensor has a sensitivity of $\frac{1}{\sqrt{2}}$ for both photon noise and read noise \cite{guyon2005limits, males2018ground}. We neglect the effects of amplitude errors due to scintillation and assume that there are no chromatic errors, as we plan to do wavefront sensing and science at the same wavelength. The results of the model are shown in Figure \ref{fig:contrast_curves}. Here we see that  a loop speed of 2 kHz provides the best contrast at small angular separations. The aniso-server error dominates at the slower loop speed of 1 kHz and at high speeds the photon noise starts to dominate which degrades contrast. Therefore, the system should run at $\sim2$ kHz for optimal performance. We simulate an optimal modal gain integral controller, following some examples in the literature\cite{gendron1995astronomical, males2018ground}.

\subsection{Improvements from predictive wavefront control methods}
Above we have investigated the required rate for our AO system and NCPA correction to achieve our contrast goal. However, we have not performed a photon budget analysis to determine if we will have enough signal-to-noise to run both of the loops at the desired rates. One strategy to allow the system to run slower while maintaining the high level of correction is to implement advanced control algorithms that optimize the control bandwidth without taking the measurement noise hit from shorter wavefront sensor exposures. A potential solution is predictive control. On-sky testing from Keck/NIRC2\cite{van2022predictive} and Subaru/SCExAO\cite{currie2019} shows that linear predictive control methods can provide a factor of $\sim2-3$ in contrast as compared to an integrator when running at 1kHz. See Fowler, 2023\cite{Fowler2023} for an overview of predictive methods. It is also possible to run the NCPA loop with a predictive step to further improve the correction~\cite{Gerard2022}. %With an arbitrarily bright guide star, the performance improvements from predictive control lessen as the control speed increases, but predictive controllers also help maintain the same performance while running the loop slower. Predictive control is likely also controlling for standard consistent vibrations and systematics, so may see some increased performance regardless. 
Perhaps more importantly however, adaptive predictive control~\cite{vanKooten19} will allow a system to constantly adapt its controller to provide optimal control \cite{haffert2021data,landman2021self, nousiainen2022toward}. This is important as the conditions during observations are constantly changing\cite{vanKooten19}.

%theoretically requires a high-speed correction to run more slowly, and increase our sky-coverage. With a 5kHz AO loop requirement, we are selecting for a very limited range of guidestars; predictive control will allow us to emulate that 5 kHz performance while still runningly slowly enough to collect a high SNR image on our wavefront sensor.  

\subsection{NCPA Correction Speed}

\begin{figure}[h!]
\makebox[\textwidth][c]{
\mbox{\includegraphics[width=0.7\textwidth]{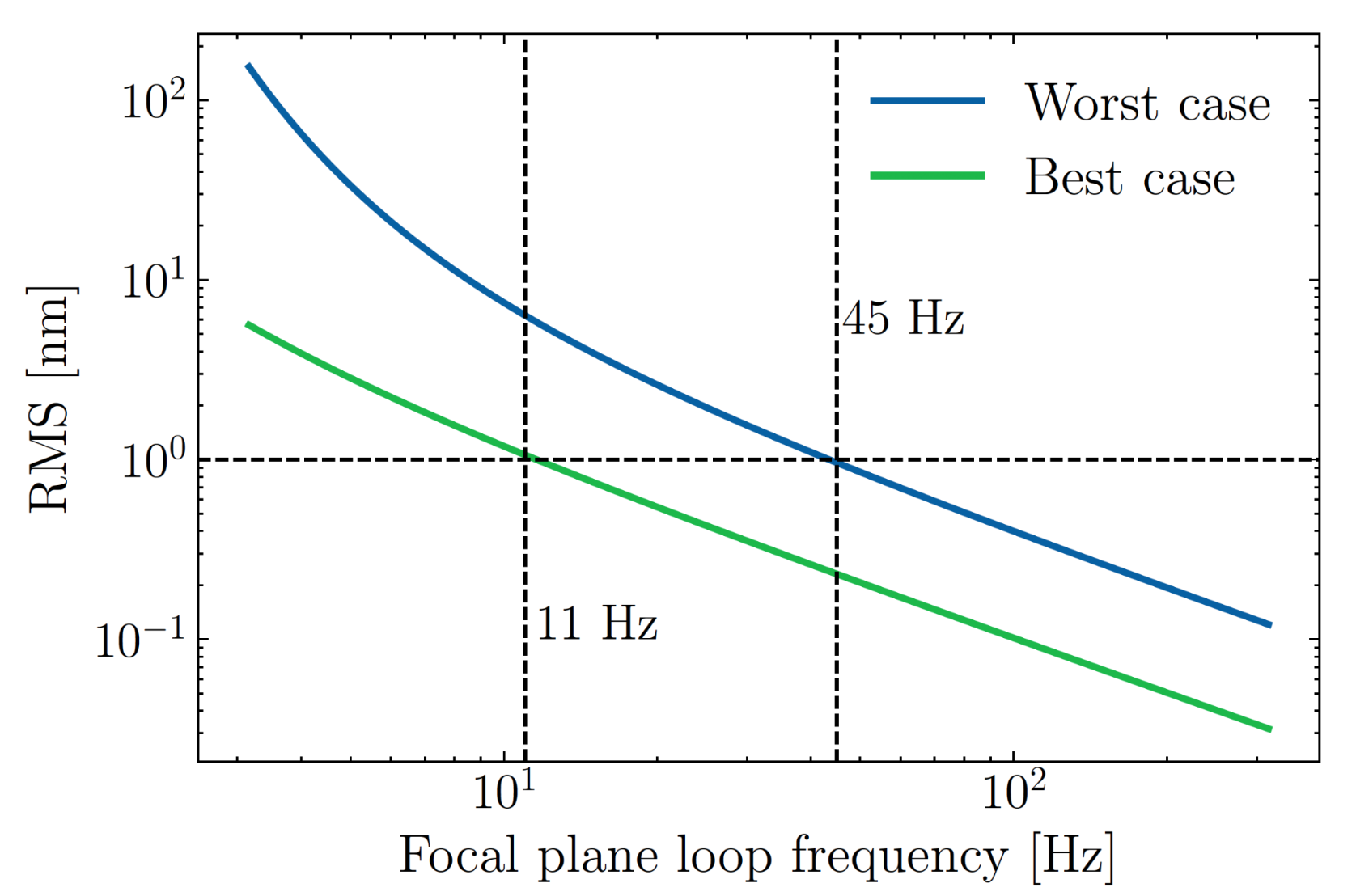}}
}
    \caption{NCPA in RMS wavefront error as a function of the focal plane control loop frequency for different scenarios. A loop speed of 11 and 45Hz is expected to maintain a wavefront error below 1nm RMS for the NCPA and thus achieve a stable contrast of $3\times 10^{-5}$.}
\label{fig:ncpa}
\end{figure}

Independent of the AO system, wavefront errors exist in the system that degrade our final performance. Specifically NCPA, whatever their origin, must be corrected to achieve good contrast. To better understand how fast we would need to run an NCPA correction, we perform a case study using an existing instrument for which the NCPA have been measured and characterized. We focus on VLT/SPHERE to determine how fast a NCPA wavefront sensor would need to run to correct for the NCPA detected by the ZELDA wavefront sensor\cite{Vigan2022}. From the NCPA decorelation equations found for SPHERE for a variety of different nights, we estimate the correction frequency of the NCPA correction loop for the best and worse case scenarios. If we want to keep the NCPAs below 1 nm RMS wavefront error, to maintain a contrast of $3\times 10^{-5}$, in the worse case we need to run the NCPA correction loop at 45 Hz as shown in Fig.~\ref{fig:ncpa}. 

\section{PROPOSED INSTRUMENT ARCHITECTURE}
\label{sec:underground}
\begin{figure}[h!]
\makebox[\textwidth][c]{
    \mbox{\includegraphics[width=1.1\textwidth]{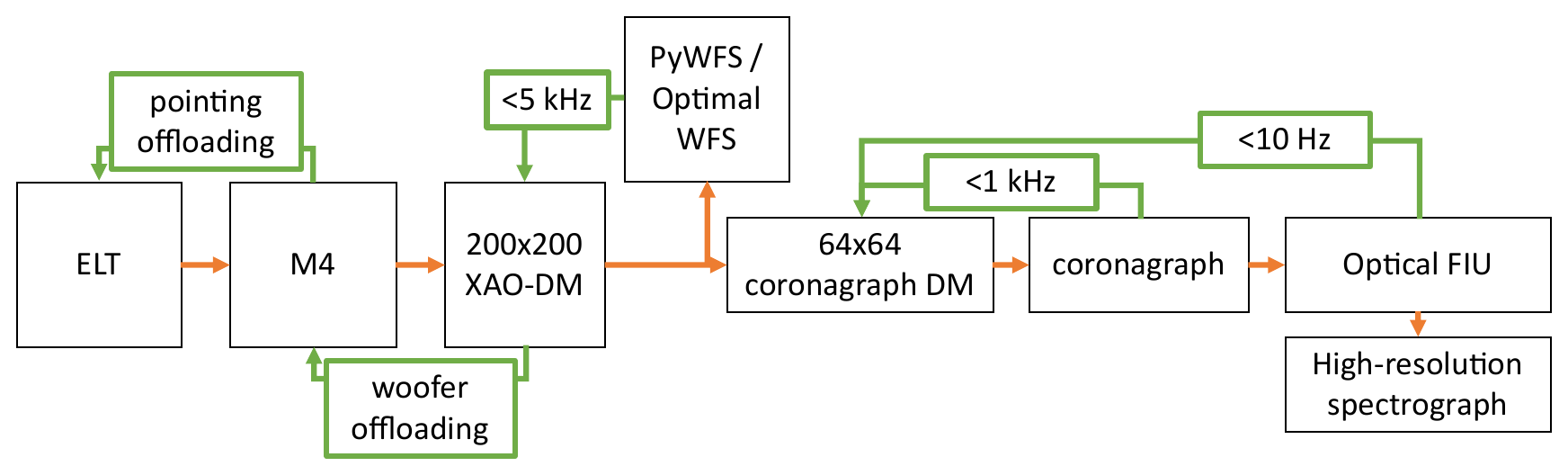}}
    }
    \caption{Schematic of the proposed UNDERGROUND instrument setup. A high-speed high-order loop corrects the atmosphere to maintain images with high Strehl ratios. A slow loop at 10 Hz creates a dark hole region where all starlight is removed. This dark hole is maintained at high speed by the wavefront sensor that is integrated into the coronagraph.}
\label{fig:optical_train}
\end{figure}

Here, we lay out our proposed instrument architecture for the UNDERGROUND instrument. We focus on 3 main technologies to mature: (1) optimal WFS at 2kHz, (2) fast NCPA control, and (3) high resolution spectroscopy.
%\label{emerging_tech}
Figure \ref{fig:optical_train} shows the proposed instrument design for the UNDERGROUND instrument. 

\textbf{Optimal WFS at 2kHz}: We need a high-speed high-order sensitive wavefront sensor that will control the exAO deformable mirror. This could be an un-modulated pyramid wavefront sensor\cite{ragazzoni2002pyramid}, the optimized Zernike Wavefront Sensor (ZWFS)\cite{chambouleyron2021}, or the optimal PIAA-ZWFS\cite{haffert2023}. Depending on the wavefront sensor, we need a different amount of pixels to sample the wavefront. The pyramid WFS  requires a detector that can run at least 440x440 pixels at 2 kHz with sub-electron read-noise. Recent advances in CMOS detector technology show much promise.\footnote{The Kinetix from Teledyne can run at 3.6 kHz in high-speed mode with 440 pixel lines albeit with 2 electrons of read noise.}

\textbf{Fast NCPA Control}: The NCPAs will be controlled by a WFS that is integrated into the coronagraph. This could be a ZWFS integrated within a Lyot-style coronagraph as demonstrated in \cite{pourcelot2022, pourcelot2023, haffert2023}. The integrated control inside the coronagraph is necessary to meet the raw contrast requirements. Additionally, we added a dedicated coronagraphic DM inside the instrument to control the focal plane speckles. This will make it easier to implement EFC or iEFC because there is no need anymore to reference offset the high-order wavefront sensor. This separates the control loops of both processes making it easier to control both. 

\textbf{High Resolution Spectroscopy}: There are several attractive solutions for high-resolution multi-object spectrographs, especially if only a narrow wavelength range is required. A promising option is the VIPA-style spectrograph\cite{zhu2020, carlotti2022sky} that can achieve high throughput and resolution in a compact design. 

%\subsection{Fast non-common path aberration control}
%    \subsubsection{Dark hole digging} (Maaike, Mamadou, Jules)
%    \begin{itemize}
%    \item Small dark zone for characterization
%    \item Short overview of techniques in one frame (SCC, PW+EFC, %    \item we want a method that does not rely on temporal diversity if we have to integrate for more than 1min  (example of SPHERE Potier's paper)
%    \end{itemize}
%    \subsubsection{Contrast maintenance} (Adrien, )
%    \begin{itemize}
%    \item Rejected/transmitted light from Lyot Stop, from FPM,     \end{itemize}

\section{Conclusion} \label{sec:conc}

Proxima Centauri b provides an exciting opportunity to detect and characterize an Earth-like planet in visible light. With high-resolution spectroscopy, observations with future Extremely Large Telescopes could detect features like oxygen, a promising biomarker that points to life on extra-solar planets. With a contrast goal of  $3\times10^{-5}$ at 10 $\lambda$/D, ground-based extreme adaptive optics combined with high-resolution multi-object spectroscopy will likely achieve this. We find that the AO system needs to exploit optimal wavefront sensing and control at $\sim2$kHz and NCPA control needs to be done at tens of Hertz. To achieve all of this we outline the UNDERGROUND instrument architecture, in Section~\ref{sec:underground}, that could enable the detection of oxygen for an Earth-like planet such as Proxima Centauri b. 
%In particular, to achieve 5kHz AO, while deformable mirrors that show promise are available, a visible-light detector that can read out a the detector area that can do wavefront sensing for a 40000 actuator DM is not available. Future development on detectors or wavefront sensing methods are required. 

\acknowledgments 
%Lorentz Center and OEI workshop.
%The Optimal Exoplanet Imaging workshop (2023) on which this manuscript is based was made possible thanks to the logistical and financial support of the Lorentz Center, Leiden, The Netherlands. This workshop was supported by Nederlandse Onderzoekschool voor Astronomie (NOVA) and by the European Research Council (ERC) under the European Union's Horizon 2020 research and innovation programme (grant agreement n°866001 - EXACT). SYH was funded by the generous support of the Heising-Simons Foundation.

The 2023 Optimal Exoplanet Imagers workshop, which sparked the
work presented in this manuscript, was made possible thanks to the
logistical and financial support of the Lorentz Center, Leiden, The
Netherlands. The research presented in this paper was initiated at
a workshop held in Leiden and partially supported by NOVA (the
Netherlands Research School for Astronomy) and by the European
Research Council (ERC) under the European Union’s Horizon 2020
research and innovation programme (grant agreement 866001 -
EXACT). SRV acknowledges funding from the European
Research Council (ERC) under the European Union’s Horizon 2020
research and innovation program under grant agreement № 805445.
SLC acknowledges support from an STFC Ernest Rutherford Fellowship. TDG acknowledges funding from the Max Planck ETH
Center for Learning Systems. KB acknowledges support from NASA
Habitable Worlds grant 80NSSC20K152, and previous support for
related work from NASA Astrobiology Institute’s Virtual Planetary
Laboratory under Cooperative Agreement NNA13AA93A. IL
acknowledges the support by a postdoctoral grant issued by the Centre National d’Études Spatiales (CNES) in France. PB, IL, and YG
were supported by the Action Spécifique Haute Résolution Angulaire
(ASHRA) of CNRS/INSU co-funded by CNES. EHP is supported by
the NASA Hubble Fellowship grant \#HST-HF2-51467.001-A awarded
by the Space Telescope Science Institute, which is operated by the
Association of Universities for Research in Astronomy, Incorporated,
under NASA contract NAS5-26555. LA and EC acknowledge funding from the European Research Council (ERC) under the European
Union’s Horizon Europe research and innovation programme (ESCAPE, grant agreement 101044152). OA and LK acknowledge
funding from the European Research Council (ERC) under the European Union’s Horizon 2020 research and innovation programme
(grant agreement № 819155). SYH was funded by the generous support of the Heising-Simons Foundation. RJP is supported by NASA
under award number 80GSFC21M0002. OHS acknowledges funding
from the Direction Scientifique Générale de l’ONERA (ARE Alioth).
This research has made use of NASA’s Astrophysics Data System Bibliographic Services and the SIMBAD database, operated at CDS,
Strasbourg, France

\bibliography{main}
\bibliographystyle{spiebib}

\end{document}